\documentclass[aps,prx,reprint,twocolumn,showpacs,floatfix,superscriptaddress]{revtex4-2}

\usepackage{amssymb,amsmath,amstext}                
\usepackage{graphicx}                                               
\usepackage{epstopdf}                                               
\usepackage{color}                                                     
\usepackage{bm}                                                        
\usepackage{appendix}                                              
\usepackage[utf8]{inputenc}
\usepackage{latexsym}
\usepackage{xcolor}
\usepackage{braket}
\usepackage{ulem}
\usepackage{siunitx}
\usepackage{umoline}
\usepackage{epsfig}
\usepackage{graphics}
\usepackage{amssymb}
\usepackage{psfrag}
\usepackage{mathtools}
\usepackage[english]{babel}
\usepackage[T1]{fontenc}
\usepackage{lmodern}
\usepackage{booktabs}

\usepackage{rotating}
\usepackage{multirow}

\definecolor{lblue} {RGB}{51,71,158}
\usepackage[colorlinks=true,citecolor=blue,linkcolor=blue,urlcolor=lblue]{hyperref}

\begin{document}

\title{Stability of many-body localization in Floquet systems}

\author{Piotr Sierant} 
\affiliation{ICFO-Institut de Ci\`encies Fot\`oniques, The Barcelona Institute of Science and Technology, Av. Carl Friedrich
Gauss 3, 08860 Castelldefels (Barcelona), Spain}
\author{Maciej Lewenstein} 
\affiliation{ICFO-Institut de Ci\`encies Fot\`oniques, The Barcelona Institute of Science and Technology, Av. Carl Friedrich
Gauss 3, 08860 Castelldefels (Barcelona), Spain}
\affiliation{ICREA, Passeig Lluis Companys 23, 08010 Barcelona, Spain}

\author{Antonello Scardicchio}
\affiliation{The Abdus Salam International Center for Theoretical Physics, Strada Costiera 11, 34151, Trieste, Italy}
\affiliation{INFN Sezione di Trieste, Via Valerio 2, 34127 Trieste, Italy}

\author{Jakub Zakrzewski} 
\affiliation{Instytut Fizyki Teoretycznej, 
Uniwersytet Jagiello\'nski,  \L{}ojasiewicza 11, PL-30-348 Krak\'ow, Poland}
\affiliation{Mark Kac Complex Systems Research Center, Uniwersytet Jagiello{\'n}ski, Krak{\'o}w, Poland}

\date{\today}

\begin{abstract}
We study many-body localization (MBL) transition in disordered Floquet systems using a polynomially filtered exact diagonalization (POLFED) algorithm. We focus on disordered kicked Ising model and quantitatively demonstrate that finite size effects at the MBL transition are less severe than in the random field XXZ spin chains widely studied in the context of MBL. Our conclusions extend also to other disordered Floquet models, indicating smaller finite size effects than those observed in the usually considered disordered autonomous spin chains.
We observe consistent signatures of the transition to MBL phase for several indicators of ergodicity breaking in the kicked Ising model. 
Moreover, we show that an assumption of a power-law divergence of the correlation length at the MBL transition yields a critical exponent $\nu \approx 2$, consistent with the Harris criterion for 1D disordered systems.
\end{abstract}

\maketitle

\section{Introduction}
 The eigenstate thermalization hypothesis \cite{Deutsch91, Srednicki94, Alessio16} predicts that 
an isolated quantum system will reach an equilibrium determined only by a few macroscopic 
conserved quantities, independently of the details of the initial state.
An exception to this ergodic paradigm is provided by a phenomenon of many-body localization (MBL) \cite{Basko06,Gornyi05,Znidaric08,Pal10, Nandkishore15, Alet18, Abanin19} which is a generic mechanism that inhibits the approach to equilibrium of interacting quantum many-body systems in the presence of disorder.
This gives rise to a dynamical phase characterized by the emergence of local 
integrals of motion \cite{Huse14,Ros15, Wahl17, Mierzejewski18, Thomson18} that preserve the information about the initial state, resulting in a suppression of transport \cite{Znidaric16, Bertini21} and a slowdown of the entanglement spreading  \cite{Chiara06, serbyn2013universal,iemini2016signatures}. Numerical studies demonstrated that finite spin-1/2 XXZ chains \cite{Oganesyan07, Santos04a, DeLuca13, Luitz15}, as well as bosonic models \cite{Sierant18, Orell19} and
systems of spinful fermions \cite{Mondaini15, Prelovsek16, Zakrzewski18, Kozarzewski18, Richter22} 
undergo MBL at sufficiently strong disorder. Also periodically driven Floquet systems may become MBL 
\cite{Lazarides15, Ponte15, Ponte15a, Abanin16, Zhang16, Bairey17, Sahay21, Garratt21, Sonner21} which allows one to avoid heating \cite{Moessner17} and enables exotic nonequilibrium phases of matter, such as time crystals \cite{Sacha15, Sacha17,Khemani16, Else16, Choi17,Bordia17, Pizzi20,Mi22} or Floquet insulators \cite{Po16, Nathan19, Roy17, Harper17, Rudner20}.
 
Recent investigations \cite{Suntajs19, Kiefer20, Sels20, Sels21a, Kiefer21, Sierant21a} of disordered many-body systems have unraveled, however, notorious difficulties in our understanding of the ergodic-to-MBL crossover. A nonmonotonic behavior of indicators of ergodicity breaking at the crossover and a limited range of system sizes (nowadays typically $L \sim 20$, restricted by the exponential growth of the Hilbert space) accessible in unbiased numerical approaches \cite{Pietracaprina18, Sierant20p, Beeumen20, Kutsuzawa22}, do not allow for an unambiguous extrapolation of the numerical results for the typically considered spin-1/2 XXZ chains to the thermodynamic limit. Consequently, it remains unclear \cite{Morningstar21, Sels21} whether the numerically observed crossover between the ergodic and MBL regimes gives rise to a MBL phase that is stable in the thermodynamic limit \cite{Imbrie16, Imbrie16a, Sierant20b, Abanin19a, Panda19, Crowley21, Ghosh22} or whether the ergodicity is restored at length and time scales that increase with the disorder strength. Notably, constrained spin chains follow the latter scenario and become ergodic in the thermodynamic limit \cite{Sierant21} despite hosting a well-pronounced MBL regime at finite system sizes \cite{Chen18pxp}.

This demonstrates the need of identifying quantum many-body systems that allow for a clearer demonstration of MBL than for the widely studied spin-1/2 XXZ chains \cite{Luitz16,Yu16, Berkelbach10, Agarwal15, Bera15, Serbyn15, Devakul15, Bertrand16,Khemani17a, Enss17,Serbyn17, Bera17, Gray18, Kjall18,Doggen18, Mace18, Herviou19,Sierant19b,Schiulaz19, Colmenarez19,Huembeli19,Chanda20m,Sierant20, TorresHerrera20, Suntajs20, Laflorencie20,Villalonga20,Villalonga20a, Vidmar21, Szoldra21, Nandy21, Kotthoff21, Hemery21}. For autonomous systems a significant step in this direction was achieved in the zero-dimensional ``quantum sun'' model \cite{Suntajs22}. In this work, we achieve this goal by performing large-scale numerical calculations for a disordered kicked Ising model (KIM) with the state-of-the-art polynomially filtered exact diagonalization (POLFED) algorithm \cite{Sierant20p, Luitz21}.
We identify ergodic, critical and MBL regimes by considering system size dependent disorder strengths $W^T_X(L)$ and $W^*_X(L)$ and quantitatively demonstrate that finite size effects at the ergodic to MBL crossover in KIM are significantly weaker than in the XXZ model. This allows us to locate the MBL transition in KIM and investigate the scenario of a power-law divergence  of correlation length at the transition. We establish robustness of our conclusions  by numerical investigations of other disordered Floquet systems.

\section{Models and methods}

\subsection{Kicked Ising model}
We consider a disordered KIM \cite{Prosen02, Prosen07} defined by the Floquet operator over one driving period for a 1D spin-1/2 chain
\begin{eqnarray}
 U_{\mathrm{KIM}}= e^{-i g \sum_{j=1}^L \sigma^x_j} e^{ -i \sum_{j=1}^L (J \sigma^z_j \sigma^z_{j+1}  +  h_j \sigma^z_j ) },
 \label{eq:KIM}
\end{eqnarray}
where $\sigma^{x,y,z}_j$ are Pauli operators, $h_j\in[0,2\pi]$ are independent, uniformly distributed random variables and periodic boundary conditions are assumed. 

We mainly focus our attention on the case $g=J=1/W$, in which $W$ plays the role of the disorder strength in the system.
The KIM is maximally ergodic for $W=4/\pi$ \cite{Akila16, Kos18, Bertini18, Lerose21}. Here, we consider higher values of $W$, up to a strong disorder limit in which $\sum_j h_j \sigma^z_j$ becomes a dominant term in $U_{\mathrm{KIM}}$.
We note that this parametrization of the system is analogous to the disordered Heisenberg spin chain \cite{Luitz15}, in which both the tunneling amplitude and the interaction strength are much smaller that the disorder amplitude in the strong disorder regime. The results for this parameter choice are presented in Sec.~\ref{sec:KIM} and in Sec.~\ref{sec:FSS}

However, our results do not depend on this particular parametrization. To demonstrate that, we consider also instances when:
\begin{itemize}
 \item the interaction strength is kept constant, $J=1$ and $g=1/W$
 \item the interaction strength is $J=1/W$ while $g=1.5$ is kept constant
 \item $g=1/W$ and the interaction is itself disordered $J=1 +  \delta J_i$, where $\delta J_i$ are independent, uniformly distributed random variables in interval $\in [-\delta J, \delta J]$ with $\delta J$ is kept constant.
\end{itemize}
In the three cases above, the parameter $W$ plays the role of the disorder strength, and the results for those Floquet systems are shown in Sec.~\ref{sec:ROB}~A.

\subsection{Other models}

To better understand the impact of symmetries and interaction range on the ergodic-MBL crossover in Floquet systems, we investigate also a family of many-body systems with Floquet operators that differ from \eqref{eq:KIM} by the operator off-diagonal in the eigenbasis of $\sigma^z_i$. We denote  $U_Z \equiv \exp [ -i \sum_{j=1}^L (J \sigma^z_j \sigma^z_{j+1}  +  h_j \sigma^z_j ) ]$ and consider models with the following Floquet operators:
\begin{eqnarray}
 U_{F,1}= e^{-i g \sum_{j=1}^L \sigma^x_j \sigma^x_{j+1}}  \, U_Z,
 \label{eq:UF1}
\end{eqnarray}
which has a $Z_2$ symmetry generated by an operator $\prod_{j=1}^L \sigma^z_j$;
\begin{eqnarray}
 U_{F,2}= e^{-i g \sum_{j=1}^L (\sigma^x_j \sigma^x_{j+1} + \sigma^y_j \sigma^y_{j+1})}  \, U_Z,
 \label{eq:UF2}
\end{eqnarray}
 which has a $U(1)$ symmetry, i.e. the total $Z$ component of the spin $\sum_{j=1}^L \sigma^z_j$ is conserved by $U_{F,2}$; 
\begin{eqnarray}
 U_{F,3}= e^{-i \frac{g}{2} \sum_{j=1}^L (\sigma^x_j + \sigma^x_j \sigma^x_{j+1})}  \, U_Z,
 \label{eq:UF3}
\end{eqnarray}
in which the interaction range is the same as in \eqref{eq:UF1}, but the model does \emph{not} have the $Z_2$ symmetry;
\begin{eqnarray}
 U_{F,4}= e^{-i \frac{g}{2} \sum_{j=1}^L (\sigma^x_j + \sigma^x_j \sigma^x_{j+1} + \frac{2}{3}\sigma^x_j \sigma^x_{j+3})}  \, U_Z,
 \label{eq:UF4}
\end{eqnarray}
which has bigger interaction range than \eqref{eq:UF3}. In all the cases above, we set $g=J=1/W$, where $W$ is the disorder strength which allows us to tune the models across the ergodic-MBL crossover. 

Finally, in order to study the role of the lack of energy conservation, we compare the ergodic-MBL crossover in the Floquet models with the results for transverse field Ising model (TFIM) with Hamiltonian given by:
\begin{eqnarray}
 H_{\mathrm{TFIM}}= \sum_{j=1}^L \sigma^x_j + \sum_{j=1}^L (J \sigma^z_j \sigma^z_{j+1}  +  h_j \sigma^z_j),
 \label{eq:TFIM}
\end{eqnarray}
where $J=1$ and $h_j$ are random variables drawn independently from interval $[-W/2, W/2]$, with $W$ denoting the disorder strength.
The results for the systems \eqref{eq:UF1}-\eqref{eq:TFIM} are shown in Sec.~\ref{sec:ROB}~B.

\subsection{Methods}
To find the eigenvectors $\ket{\psi_n}$ and the corresponding eigenvalues $e^{i \phi_n }$ of the Floquet operators $U_{\mathrm{KIM}}$ and $U_{F,k}$ for $k=1,3,4$, we use the POLFED algorithm \cite{Sierant20p} employing a geometric sum filtering \cite{Luitz21}. The performance of the algorithm relies crucially on the efficiency of matrix vector multiplication, with matrix being one of the respective Floquet operators. The Floquet operators $U_{\mathrm{KIM}}$ and $U_{F,k}$ with $k=1,3,4$, are products of operators that are diagonal in the eigenbases of $\sigma^x_i$ and $\sigma^z_i$ operators. Hence, the matrix vector multiplication can be performed efficiently by switching between the two bases by means of a fast Hadamard transform \cite{Fino76, Arndt11}, and acting with the appropriate diagonal matrix (see App.~\ref{app:POLFED} for details).
This allows us to obtain eigenstates $\ket{\psi_n}$ for system sizes $L \leq 20$, significantly larger than for $L \leq 14$ considered in earlier exact diagonalization studies of KIM \cite{Zhang16, Sonner21}.
The $U(1)$ symmetric Floquet operator $U_{F,2}$ \eqref{eq:UF2} is diagonal in the momentum basis rather than in the eigenbasis of $\sigma^x_i$. Therefore, we investigate  $U_{F,2}$ by means of a full exact diagonalization, reaching system sizes up to $L=16$. Finally, to find eigenstates in the middle of the spectrum of TFIM \eqref{eq:TFIM}, we directly employ the POLFED algorithm for Hermitian matrices as described in \cite{Sierant20p}.

 \begin{figure*}
 \includegraphics[width=0.99\linewidth]{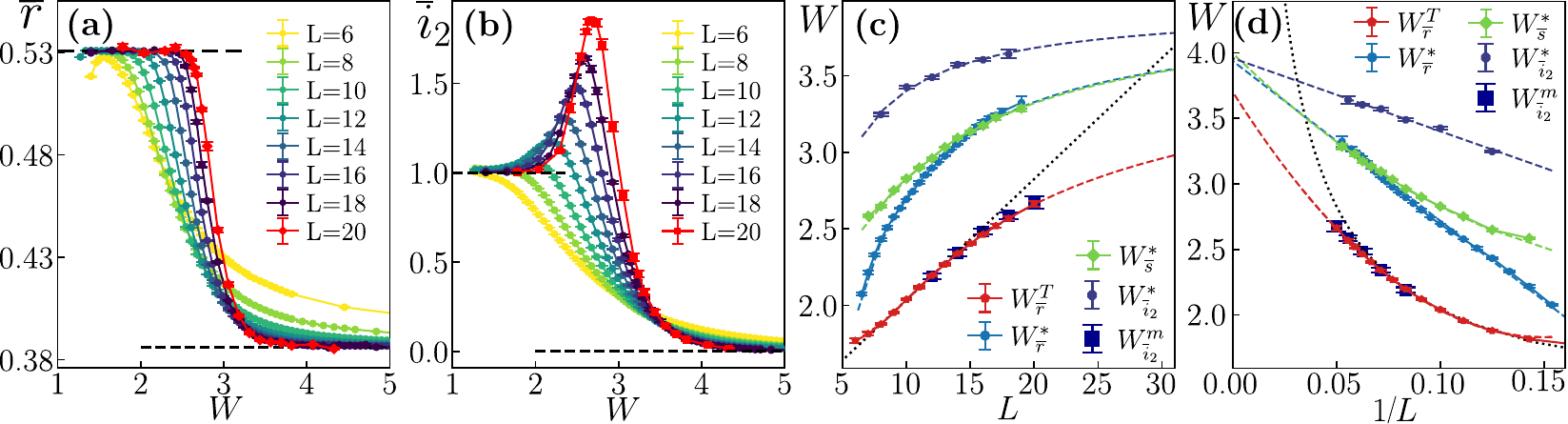}
  \caption{The ergodic-MBL crossover in KIM \eqref{eq:KIM} with $g=J=1/W$. 
  Gap ratio $\overline r$ \textbf{(a)} and rescaled QMI $\overline{i}_2$ \textbf{(b)} as function of disorder strength $W$ for system size $L$; dashed lines correspond to predictions for ergodic and MBL systems. Disorder strength  $W^T_{\overline r}$ at which $\overline r$ departs from the ergodic value and the crossing points $W^*_X$ as function of $L$ \textbf{(c)} and $1/L$ \textbf{(d)} where $X$ is either the gap ratio $\overline r$, the rescaled entanglement entropy $ \overline s$, or the rescaled QMI $\overline{i}_2$; the dotted lines denote $W_T(L) \sim L$ scaling; the dashed lines correspond to fits $W(L)=W_{\infty}+a/L+b/L^2$ with $W_{\infty}=3.97\pm0.03$ for $W^*_{\overline r}(L)$, $W^*_{\overline s}(L)$ and $W^*_{\overline{i}_2}(L)$.
 }\label{figCOMB1}
\end{figure*}  
     
\section{Results for kicked Ising model}
\label{sec:KIM}
     
In this section, we investigate a crossover between ergodic and MBL regimes in KIM at finite system size $L$. Throughout this section, we set $g=J=1/W$. By introducing system size dependent disorder strengths $W^T_{X}(L)$ and $W^*_X(L)$, we show that finite size effects at the MBL crossover in KIM are better controlled than in the disordered XXZ spin chain.

\subsection{Ergodic-MBL crossover in KIM}
We calculate $N_{\mathrm{ev}} = \mathrm{min}\{ 2^L /10, 1000\}$ eigenvectors $\ket{\psi_n}$ of $U_{\mathrm{KIM}}$. Due to the constant density of eigenphases $\phi_n$, we can treat each eigenvector on equal footing. For concreteness, we choose eigenstates with eigenphases $\phi_n$ closest to $0$ and average results over more than $5\cdot 10^4$, $5\cdot 10^3$ and $5\cdot 10^2$ disorder realizations, respectively for $L\leq16$, $L = 17, 18$ and $L=20$, see App.~\ref{app:error} for analysis of statistical errors.  

To probe the properties of eigenphases, we compute the gap ratio 
\begin{equation}
\overline r= \left \langle \min \{g_{i},g_{i+1} \} / \max\{g_{i},g_{i+1}\} \right \rangle 
\end{equation}
 where $g_i=\phi_{i+1}-\phi_{i}$ and $\left \langle . \right \rangle$ denotes the average over the calculated fraction of spectrum and disorder realizations. 
We study also the entanglement of eigenstates $\ket{\psi_n}$.
The entanglement entropy \cite{Amico08} is given by 
\begin{equation}
S(A)=-\sum_{i=1}^{i_M} \alpha_{i}^2 \log(\alpha_{i}^2), 
\end{equation}
where $\alpha_{i+1}>\alpha_i$ are Schmidt basis coefficients \cite{Karol} of the eigenstate $\ket{\psi_n}$
for a partition of the 1D lattice into a subsystem $A$ and its complement.
Choosing $A=[1, L/2]$, we calculate the rescaled entanglement entropy $\overline s = \left \langle S(A) \right \rangle/S_{COE}$ 
by taking the average $\left \langle . \right \rangle$ over the eigenstates, disorder realizations and rescaling the result by numerically calculated average entanglement entropy $S_{COE}$ of eigenstates of Circular Orthogonal Ensemble of random matrices (COE) that models the properties of $U_{\mathrm{KIM}}$ in the ergodic regime \cite{Alessio14, Vidmar17}.
For $A=[1, L/2]$ we also calculate the average Schmidt gap $\Delta = \left \langle \alpha_{1}^2 - \alpha_{2}^2\right \rangle$. Furthermore, we calculate the quantum mutual information (QMI) $I_2 = S(B)+S(C)-S(B\cup C)$ for the subsystems $B=[1, \left \lceil{ L/4}\right \rceil ]$, $C=(2\left  \lceil{ L/4}\right\rceil, 2\left  \lceil{ L/4}\right\rceil+\left \lfloor{ L/4}\right \rfloor ]$ (where $\left  \lceil{ .}\right\rceil$, $\left \lfloor{ . }\right \rfloor$ denote the ceiling and floor functions), and obtain the rescaled QMI as
$\overline i_2 = \left \langle I_2 \right \rangle / I_{COE}$ where $I_{COE}$ is the average QMI for COE eigenstates. Also, we
 compute the spin stiffness $\overline C = \left \langle \sum_i || \bra{\psi_n} \sigma^z_i \ket{\psi_n} ||^2 \right \rangle /L$ which is an infinite time average of the spin-spin autocorrelation function  $C(t)=\sum_i \mathrm{Tr}[\sigma^z_i(t)\sigma^z_i(0)]/( L 2^L )$.

As the strength of the disorder, $W$, increases, the gap ratio $\overline r$, shown in Fig.~\ref{figCOMB1}(a), decreases from
$\overline r = \overline r_{COE}\approx 0.53$ characteristic for the ergodic regime
to $\overline r = \overline r_{PS}\approx 0.386$ for an MBL system \cite{Atas13}. The QMI \cite{Groisman05} measures the total amount of correlations between the subsystems $B$, $C$ and decays exponentially with the distance between the subsystems in the MBL regime \cite{Tomasi17}. In the ergodic regime, the volume-law terms proportional to the lengths of subsystems $B$, $C$, $B \cup C$ cancel out and the QMI is equal to a system size independent value $I_{COE}$. Consequently, in Fig.~\ref{figCOMB1}(b), we observe a crossover in the rescaled QMI $\overline i_2$ as a function of $W$ between the limiting values $\overline i_2 = 1$ and $\overline i_2=0$. The correlations between the subsystems are enhanced at the crossover, hence the rescaled QMI admits a maximum between the ergodic and MBL regimes. We observe the ergodic-MBL  crossover also in the behavior of the rescaled entanglement entropy $\overline s$, Schmidt gap $\overline \Delta$, and spin stiffness $\overline C$, see App.~\ref{app:addNUMkim}.

\subsection{Finite size effects at the MBL crossover }

To investigate the ergodic-MBL crossover we consider two 
system-size dependent disorder strengths:
\begin{enumerate}
 \item $W^T_{X}(L)$ -- the disorder strength for which, at a given system size $L$, the quantity $X$ is deviates by a small parameter $p_X$ from its ergodic value
 \item $W^*_X(L)$ -- the disorder strength for which the curves $X(W)$
cross for the system sizes $L-\Delta L$ and $L+\Delta L$ (where $\Delta L \ll L$).
\end{enumerate}
The disorder strengths $W^T_{X}(L)$ and $W^*_X(L)$ allow us to analyze the ergodic-MBL crossover in a quantitative fashion without resorting to any model of the transition. This is particularly advantageous in view of the recent controversies around the MBL transition \cite{Suntajs19, Kiefer20, Sels20, Sels21a, Kiefer21, Sierant21a, Sierant20b, Abanin19a, Panda19, Crowley21, Ghosh22}. 
The disorder strength $W^T_{X}(L)$ may be considered as a boundary of the ergodic regime, whereas $W^*_X(L)$ provides an estimate of the critical disorder strength  at given $L$. A regime between $W^T_{X}(L)$ and $W^*_X(L)$ is a critical region, vanishing for $L\to\infty$ if a transition between ergodic and MBL phases indeed occurs.

For the disordered XXZ model, both disorder strengths increase monotonously with system size: 
 $W^T_{X}(L)\sim L$ and $W^*_{X}(L)\sim W_C- \mathrm{const}/L$ for $X=\overline r, \overline s \,\,$\cite{Sierant20p}.
The latter scaling suggests a finite critical disorder strength $W_C\approx5.4$
(larger than $W_C\approx 3.7$ \cite{Luitz15}, but consistent with \cite{Devakul15, Gray18}).
However, the scalings of $W^T_{X}(L)$ and $W^*_{X}(L)$ are incompatible in the large system size limit, as  $W^T_{X}(L)$ exceeds $W^*_{X}(L)$ at $L \geq L^{\mathrm{XXZ} }_0\approx 50$ whereas
$W^T_{X}(L) < W^*_{X}(L)$ by construction at any $L$.
Therefore, when approaching the length scale $L^{\mathrm{XXZ} }_0$ (which appeared also in \cite{Panda19, Suntajs20}) one of the scalings must break down indicating either a presence of the MBL phase in the thermodynamic limit at $W>W_C$ (where $W_C \geq W^{T,*}_{X}(L)$) or showing the absence of the MBL phase (for example, when the linear increase of $W^T_{X}(L)$ prevails). However, the length scale $L^{\mathrm{XXZ} }_0$ is far beyond the reach of present day exact numerical calculations for the XXZ spin chain, which prevents one from unambiguously deciding which of the scenarios is realized in that model. Interestingly, numerical calculations for considerably larger system sizes of constrained spin chains suggest the second scenario: $W^T_{X}(L)\sim L$, $\,W^*_{X}(L) \sim L$ in which the extent of the ergodic regime increases indefinitely in the thermodynamic limit \cite{Sierant21}.

 For the investigated KIM, we start by considering the gap ratio $X=\overline r$ and we set from now on $p_{\overline r}=0.01$ (unless otherwise noted), which yields $W^T_{\overline r}(L)$ shown in Fig.~\ref{figCOMB1}(c)-(d). We observe a linear scaling $W^T_{\overline r}(L)\sim L$ with system size $L$ for $8\leq L \leq14$. Importantly, in contrast to the persistent linear drift of $W^T_{\overline r}(L) \sim L$ for XXZ spin chains, we see a clear deviation from the linear scaling for $L \geq 15$ for KIM. Therefore, the growth of $W^T_{\overline r}(L)$ with $L$ is sublinear at sufficiently large system sizes, which is a first premise suggesting the stability of MBL in KIM in the $L\to \infty$ limit. 
Accessing system sizes $15 \leq L \leq 20$ with POLFED was necessary to uncover this premise for the MBL phase in KIM.
The scaling of $W^T_{ \overline r}(L)$ remains quantitatively the same for $0.002 < p_{\overline r}<0.03$ and $W^T_{ \overline s}(L)$ behaves analogously, see App.~\ref{app:px}. The disorder strength $W^T_{\overline r}(L)$ at which the gap ratio deviates from its ergodic value $\overline r_{COE}$ coincides, to a good approximation, with the maximum $W^m_{\overline i_2}(L)$ of the rescaled QMI $\overline i_2$ which becomes pronounced at $L \geq12$ (cf. Fig.~\ref{figCOMB1}(b)). The point $W^m_{\overline i_2}$ of the maximal correlations between subsystems $B$ and $C$ follows the linear scaling of $W^T_{\overline r}(L)$ for $L=12,14$ and deviates from it at $L\geq 16$.

Now, we turn to examination of the crossing point $W^*_X(L)$ in KIM. We use 
$|L_1$$-$$L_2|$$\leq\,$$\, 2$ for $X\,$$ =$$\,\overline r$, $|L_1$$-$$L_2|\,$$=$$\,2$ for $X\,$$=$$\, \overline s$ and $|L_1$$-$$L_2|\,$$=$$\,4$ for $X\,$$=$$\,\overline i_2$ and obtain $W^*_{\overline r}(L)$, $W^*_{\overline s}(L)$, $W^*_{\overline i_2}(L)$ shown in Fig.~\ref{figCOMB1}(c)-(d). The crossing points $W^*_{\overline r}(L)$ and $W^*_{\overline s}(L)$ differ at $L \lesssim 12$, but approach each other as the size of the system increases.
Both $W^*_{\overline r}(L)$ and $W^*_{\overline s}(L)$ are well fitted by a second-order polynomial in $1/L$ whose extrapolation crosses with the extrapolation of the \textit{linear} scaling of $W^*_{\overline r}(L)$ at $L^{\mathrm{KIM}}_0\approx 28$. The length scale $L^{\mathrm{KIM}}_0\approx 28$ is significantly smaller than the analogous length scale $L^{\mathrm{XXZ} }_0\approx 50$ for XXZ spin chain. Therefore, the maximal system size investigated for KIM 
relative to this length scale, $L/L^{\mathrm{KIM}}_0 \approx 0.71$, is considerably larger than for the XXZ model $L/L^{\mathrm{XXZ}}_0 \approx 0.44$ \cite{footnoteL}.
This is the basis of a second premise that the ergodic-MBL crossover observed in KIM is stable in the large $L$ limit. The crossing points $W^*_{\overline i_2}(L)$ of the rescaled QMI $\,\overline i_2$ lie considerably above $W^*_{\overline r}(L)$, $W^*_{\overline s}(L)$. However, as shown in  Fig.~\ref{figCOMB1}(d), $W^*_{\overline i_2}(L)$ is well fitted by a first order polynomial in $1/L$. An extrapolation of this polynomial to $L\to \infty$ limit gives a result consistent with extrapolations for $W^*_{\overline r}(L)$ and $W^*_{\overline s}(L)$, suggesting that the rescaled QMI $\overline i_2$ is subject to weaker finite size effects than $\overline r$ or $\overline s$ (cf. \cite{Zabalo20, Sierant21b}).  The extrapolations yield an estimate of the critical disorder strength $W_{\infty}=3.97\pm0.03$ \cite{footnoteWT}.

In conclusion, our results for KIM indicate that finite size effects at the MBL crossover are significantly weaker than in the disordered XXZ model. 
The first premise suggesting the occurrence of MBL transition in the model
is the deviation from the linear scaling $W^T_{X}(L) \sim L$ to a weaker system size dependence. The length scale $L^{\mathrm{KIM}}_0$ characterizing the ergodic-MBL is significantly smaller in KIM than the corresponding length scale in the XXZ model. This, together with the fact that the extrapolations of the crossing points $W^*_{\overline r}(L)$, $W^*_{\overline s}(L)$ and $W^*_{\overline{i}_2}(L)$ yield consistent value of $W_{\infty}$ is the second premise for the occurrence of MBL transition in KIM.



\section{Robustness of the results}
\label{sec:ROB}
In this section we demonstrate that the conclusions of the preceding section 
apply also for other parameter choices in KIM as well as in different disordered Floquet systems. Considering various systems, we exhibit the role of symmetries and interaction range on finite size effects at the ergodic-MBL crossover. 

\begin{figure}
 \includegraphics[width=0.99\linewidth]{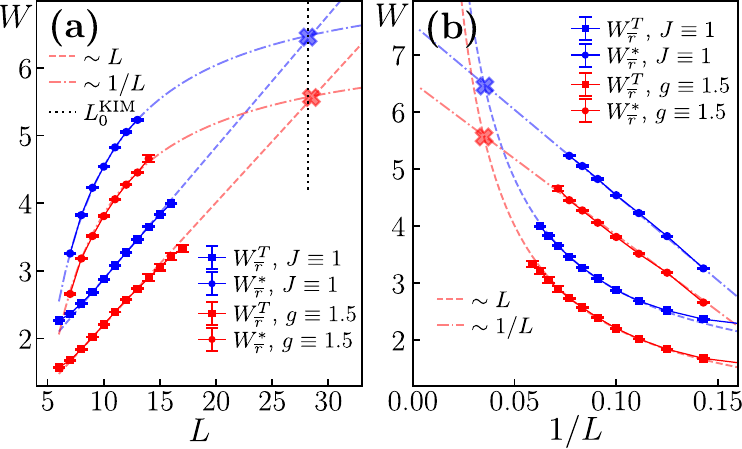} 
  \caption{The ergodic-MBL crossover in KIM \eqref{eq:KIM} with $J$$=$$1$, $g$$=$$1/W$ (denoted by blue) or $g$$=$$1.5$ and $J$$=$$1/W$ (denoted by red). Disorder strength  $W^T_{\overline r}$ at which $\overline r$ departs from the ergodic value and the crossing points $W^*_{\overline r}$ as function of $L$ \textbf{(a)} and $1/L$ \textbf{(b)}; the dashed lines denote $W_T(L) \sim L$ scaling; the dash-dotted lines correspond to fits $W(L)=W_{\infty}+a/L$. The crosses denote the length scales $L^{J}_0$ and $L^{g}_0$, whereas $L^{\mathrm{KIM}}_0$ is denoted by the vertical dotted line.
 }\label{fig:JGfix}
\end{figure}

\begin{figure}
 \includegraphics[width=0.99\linewidth]{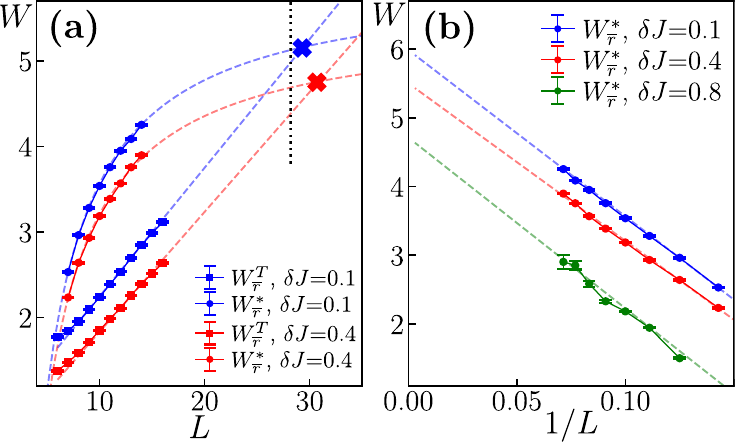} 
  \caption{The ergodic-MBL crossover in KIM \eqref{eq:KIM} with $g$$=$$1/W$ and disordered interaction term $J=1+\delta J_i$; the data for $\delta J_i=0.1, 0.4, 0.8$ are respectively denoted by blue, red and green. Disorder strength  $W^T_{\overline r}$ at which $\overline r$ departs from the GOE value and the crossing points $W^*_{\overline r}$ as function of $L$ \textbf{(a)} and $1/L$ \textbf{(b)}; the dashed lines denote $W_T(L) \sim L$ scaling; the dash-dotted lines correspond to fits $W(L)=W_{\infty}+a/L$. The crosses denote the length scale $L^{\delta J}_0$, whereas $L^{\mathrm{KIM}}_0$ is denoted by the vertical dotted line.
 }\label{fig:Jdis}
\end{figure}  
\subsection{KIM -- different parameter choices}

In Sec.~\ref{sec:KIM}, we set $g=J=1/W$. This means that both the interaction term $\sum_j J\sigma^z_j \sigma^z_{j+1}$ as well as the off-diagonal term $g\sum_j \sigma^x_j$ vanish in the strong disorder limit $W\to \infty$.

One possible choice of the parameters is to fix the interaction strength as $J$$=$$1$ and vary $g$$=$$1/W$, where $W$ is the amplitude of the disorder in the system. Another option is to fix the value of $g$, for instance choosing $g=1.5$, and vary $J=1/W$. Performing numerical calculations for both cases, we find a crossover between ergodic and MBL regimes as a function of $W$, qualitatively similar to the one shown in Fig.~\ref{figCOMB1}~(a) and (b). We reach system sizes up to $L=17$ and average the results over no less than $10^4$ disorder realizations.

Focusing, for simplicity, on the average gap ratio $\overline r$, we extract the disorder strengths $W^T_{\overline r}$ and $W^*_{\overline r}$, shown in Fig.~\ref{fig:JGfix}. The results are quantitatively similar to the ones reported in Sec.~\ref{sec:KIM}. We find a deviation from the linear scaling $W^T_{\overline r} \sim L$ to a weaker system size dependence at the largest available system sizes. Moreover, the crossing point $W^*_{\overline r}$ is well described by a first order polynomial in $1/L$. The extrapolation of this behavior crosses with the extrapolation of the linear scaling $W^T_{\overline r} \sim L$ in both models at $L^{J}_0 \approx L^{g}_0 \approx 28$, analogously to the results for KIM with $g=J=1/W$.

Now, we consider a situation in which $g=1/W$ and the interaction term is disordered, $J=1+\delta J_i$. In that case $W$ plays the role of the disorder strength which allows us to tune the system across the ergodic-MBL crossover. At the same time, the amplitude $\delta J_i$ of the disorder in the interaction term is kept fixed. The results for $\delta J =0.1, 0.4$, shown in Fig.~\ref{fig:Jdis} are, again, fully analogous to that obtained  for KIM. In particular, the system size $L^{\delta J}_0$, at which the extrapolation of the linear behavior $W^T_{\overline r} \sim L$ crosses the extrapolation of $W^*_{\overline r}$, is very close to $L^{\mathrm{KIM}}_0$ both for $\delta J=0.1$ and $\delta J=0.4$. In turn, for a sufficiently large value of $\delta J$ (e.g. $\delta J=0.8$), the GOE value is \emph{not} reached by the average gap ratio $\overline r$ at small values of $W$ (although we observe that $\overline r$ increases with system size $L$). Therefore, for $\delta J=0.8$, we extract only the position of the crossing point $W^*_{\overline r}$ which is well approximated by a first order polynomial in $1/L$, similarly to the all other cases discussed. We note that the presence of Ising-even disorder, $\delta J >0$ is necessary  for a stabilization of Floquet time crystals \cite{Ippoliti21}.

The results of this section illustrate that the finite size trends at the ergodic-MBL crossover reported in Sec.~\ref{sec:KIM} are robust to changes in the model such as fixing $g$ or $J$ or introducing a certain amount of disorder into the interaction term. In the following section we study the impact of symmetries or of the increase of the interaction range on the finite size effects at the MBL crossover.

\begin{figure}
 \includegraphics[width=0.99\linewidth]{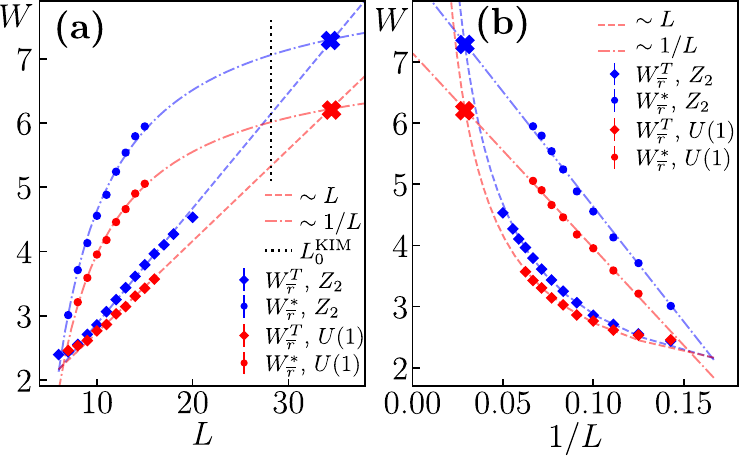} 
  \caption{The ergodic-MBL crossover in Floquet models with $Z_2$ and $U(1)$ symmetry. Disorder strength  $W^T_{\overline r}$ at which $\overline r$ departs from the ergodic value and the crossing points $W^*_{\overline r}$ as function of $L$ \textbf{(a)} and $1/L$ \textbf{(b)}; the dashed lines denote $W_T(L) \sim L$ scaling; the dash-dotted lines correspond to fits $W(L)=W_{\infty}+a/L$. The crosses denote the length scales $L^{Z_2}_0$ and $L^{U(1)}_0$, for comparison, the vertical dotted line corresponds to $L^{\mathrm{KIM}}_0$.
 }\label{fig:sym}
\end{figure}

\subsection{Other disordered models}

There are significant differences in finite size effects, reflected by the length scale $L_0$ between the KIM studied in Sec.~\ref{sec:KIM} and the disordered XXZ spin chain widely considered as a paradigmatic model of MBL.
There are two major differences between these two models that may be responsible for this disparity. 
The first difference is the fact that KIM, in contrast to the XXZ spin chain, is a Floquet model that does not conserve the energy. The second difference is the fact that the XXZ spin chain possesses the $U(1)$ symmetry associated with conservation of $\sum_i \sigma^z_i$, whereas KIM does not. 

In order to investigate the role of the abelian symmetries on ergodic-MBL crossover, we consider modifications of KIM that possess the $Z_2$ symmetry \eqref{eq:UF1} and the $U(1)$ symmetry \eqref{eq:UF2}. We extract the disorder strengths $W^T_{\overline r}(L)$ and $W^*_{\overline r}(L)$. The results, shown in Fig.~\ref{fig:sym}, show that the system size dependencies in $W^T_{\overline r}(L)$ and $W^*_{\overline r}(L)$ are analogous to KIM. We find the length scales $L^{Z_2}_0 \approx L^{U(1)}_0 \approx 34.5$ which is considerably larger than $L^{\mathrm{KIM}}_0$.

\begin{figure}
 \includegraphics[width=0.99\linewidth]{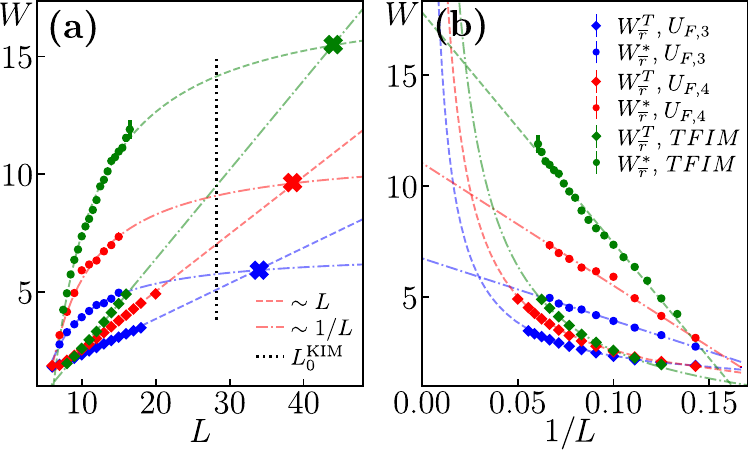} 
  \caption{
  The ergodic-MBL crossover in Floquet models \eqref{eq:UF3} (data in blue, denoted by $U_{F,3}$), \eqref{eq:UF4} (data in red, denoted by $U_{F,4}$) and in the TFIM \eqref{eq:TFIM} (data for TFIM are plotted in green and the disorder strength is transformed linearly, according to $W \to 2W$$-$$5.5$,
  for clarity of the plot). Disorder strength  $W^T_{\overline r}$ at which $\overline r$ departs from the ergodic value and the crossing points $W^*_{\overline r}$ as function of $L$ \textbf{(a)} and $1/L$ \textbf{(b)}; the dashed lines denote $W_T(L) \sim L$ scaling; the dash-dotted lines correspond to fits $W(L)=W_{\infty}+a/L$. The crosses denote the length scales $L^{U_{F,3}}_0$, $L^{U_{F,4}}_0$, and $L^{\mathrm{TFIM}}_0$; for comparison, the vertical dotted line corresponds to $L^{\mathrm{KIM}}_0$.
 }\label{fig:NOsym}
\end{figure} 

At the first sight, those results could suggest that the absence of the $Z_2$ and $U(1)$ symmetries enhances the MBL regime in the KIM. This is, however, \emph{not} the case. To demonstrate this, we consider the Floquet model \eqref{eq:UF3}, which is not $Z_2$ symmetric due to the presence of the $ \sum_j \sigma^x_j$ term. Additionally, the off-diagonal part of $U_{F,3}$ in eigenbasis of $\sigma^z_i$ contains terms coupling at most the neighboring sites of the lattice. In that sense, the interaction range of $U_{F,3}$ is the same as of the $Z_2$ symmetric Floquet operator $U_{F,1}$. The behavior of $W^T_{\overline r}(L)$ and $W^*_{\overline r}(L)$ for the $U_{F,3}$ model is shown in Fig.~\ref{fig:NOsym}. The resulting length scale $L^{U_{F,3}}_0 \approx 34$ is nearly the same as $L^{Z_2}_0$. This shows that it is the interaction range, rather than presence of the $Z_2$ symmetry that influences the length scale $L_0$ and has significant impact on finite-size effects at the MBL crossover. To confirm this hypothesis, we consider $U_{F,4}$, given by \eqref{eq:UF4}, which has an additional term $\sum_j \sigma^x_{j}\sigma^x_{j+3}$ that couples spins separated by two sites. The presence of this term increases the characteristic length scale to $L^{U_{F,4}}_0 \approx 38.5$, showing, in agreement with intuitive expectations, that an increase of the interaction range makes the finite size effects at the MBL crossover more severe.

The results so far indicate that the presence of abelian symmetries such as $Z_2$ or $U(1)$ does not have a significant effect on the finite size effects at the MBL crossover. From the perspective of the above results, part of the difference between KIM and the disordered XXZ model may stem from the bigger range of the hopping term in the latter model. The small dissimilarity between  $L^{\mathrm{KIM}}_0$ and $L^{U_{F,3}}_0$ suggests, however, that the latter factor plays a minor role. This, in turn, suggests that the energy conservation, which is the remaining disparity between the two models, has a major impact on the finite size effects at the MBL crossover.

To show that this is indeed the case, we calculate the average gap ratio $\overline r$ for TFIM \eqref{eq:TFIM}, averaging results over $N'_{\mathrm{ev}} = \mathrm{min}\{ 2^L /20, 1000\}$ 
eigenvalues in the middle of the spectrum and over no less than  $5\cdot 10^4$ ($5\cdot 10^3$) disorder realizations for $L $$\leq $$16$ ($L$$=$$17$). Extracting $W^T_{\overline r}(L)$ and $W^*_{\overline r}(L)$, we find the characteristic length scale $L^{\mathrm{TFIM}}_0 \approx 44$, see Fig.~\ref{fig:NOsym}. This length scale is significantly larger than $L^{\mathrm{KIM}}_0$, even though the terms used to construct the Hamiltonian of TFIM and the Floquet operator of KIM are the same (and thus have the same range). Thus, we conclude that the difference between KIM and disordered XXZ model that plays the major role in the finite size effects at the MBL crossover is the lack of energy conservation of the former model.

\begin{figure*}
 \includegraphics[width=0.99\linewidth]{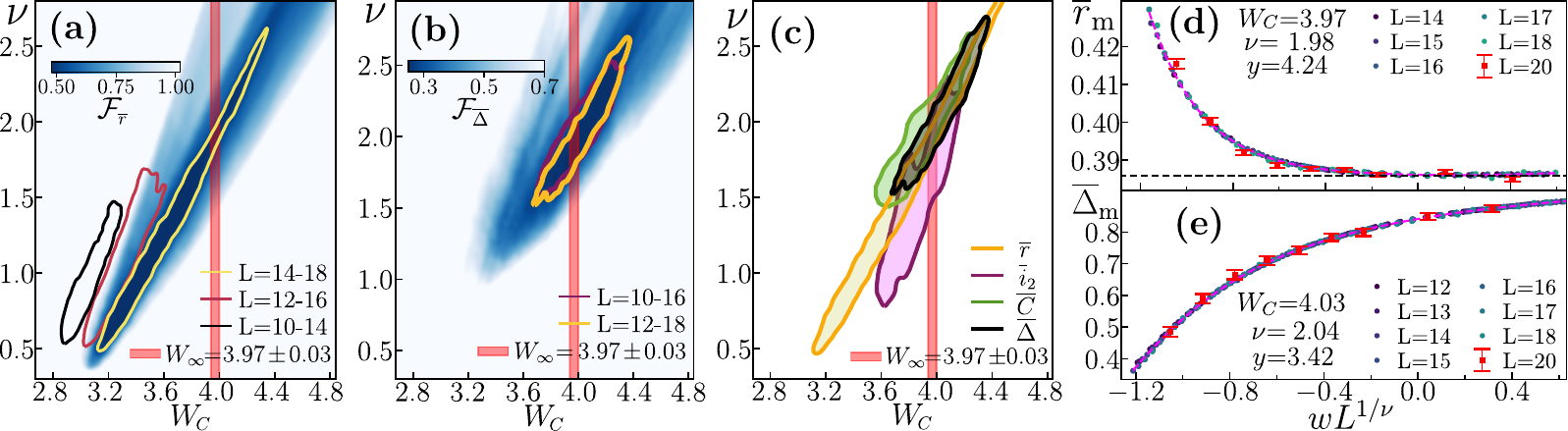} 
  \caption{ Finite size scaling analysis of ergodic-MBL crossover in KIM \eqref{eq:KIM} for $g=J=1/W$. Cost functions $\mathcal F_{\overline r}$ \textbf{(a)} and $\mathcal F_{\overline \Delta}$ \textbf{(b)} are color coded for fixed $\nu$, $W_C$, respectively for $\overline r$ (system sizes considered in the collapse $L=14$$-$$18$) and $\overline \Delta$ (for $L=12$$-$$18$).
  The contours encompass $\nu$, $W_C$ for which $\mathcal F_{X}$
  is smaller than $\frac{4}{3}$ of its minimum $\mathcal F^{\mathrm{min}}_{X}$. 
  \textbf{(c)}: the contours $\mathcal F_{X} = \mu_X \mathcal F^{\mathrm{min}}_{X}$ for collapses of gap ratio $\overline r$ ($L=14$$-$$18$, for $\overline r \leq 0.43$),
  the Schmidt gap $\overline \Delta$ ($L=12$$-$$18$, for  $\overline \Delta \geq 0.44$),
  the rescaled QMI $\overline i_2$ ($L=14$$-$$20$, for  $\overline i_2 \leq 0.3$), 
  the spin stiffness $\overline C$ ($L=12$$-$$18$, for $\overline C \geq 0.3$),
  $\mu_X=\frac{4}{3}$ for $X=\overline r, \overline \Delta, \overline C$ and $\mu_{\overline i_2}=2$.
 Collapses for $\overline r_{\mathrm{m}}$, $\overline \Delta_{\mathrm{m}}$ shown in \textbf{(d)}, \textbf{(e)}.
 }\label{figCOMB2}
\end{figure*}  
\section{Finite-size scaling analysis for MBL in kicked Ising model}
\label{sec:FSS}
We now turn to finite-size scaling (FSS) analysis  of the ergodic-MBL crossover in KIM, assumming that $g=J=1/W$, similarly as in Sec.~\ref{sec:KIM}. The MBL transition in XXZ spin chains was analyzed in the framework of power-law divergence of the correlation length \cite{Kjall14, Luitz15, Khemani17} and of Kostelitz-Thouless-like scaling \cite{Suntajs20, Laflorencie20, Hopjan21, Aramthottil21} suggested by an avalanche mechanism of thermalization \cite{DeRoeck17, Luitz17}. Both scenarios were considered within the phenomenological renormalization group approaches \cite{Vosk15, Potter15,Goremykina19,Dumitrescu19,Morningstar20}. Restricting the FSS to the vicinity of the critical disorder strength, which seems to be necessary, as exemplified by investigations of the 3D Anderson model \cite{Suntajs21}, we cannot determine which of the scenarios of the MBL transition is realized in KIM. In the following, we assume the power-law divergence of the correlation length. Investigations of Anderson transition \cite{Slevin99,Ueoka14, Tarquini17,Slevin18, Pino20} suggest then the FSS ansatz:
\begin{eqnarray}
X(W,L)=\psi_0(wL^{1/\nu})+L^{-y}\psi_1(wL^{1/\nu}),
\label{eqScal}
\end{eqnarray}
where $X$ is the quantity analyzed, $w=(W-W_C)/W_C$ is the dimensionless distance from the critical point $W_C$, $\nu$ is the exponent describing the divergence of correlation length and the exponent $y$ takes into account the corrections to the scaling due to irrelevant variables. We use the parametrization $\psi_1(wL^{1/\nu})=a_0+a_1 wL^{1/\nu}$, and consider the variable $X_{\mathrm{m}}\equiv X-L^{-y}\psi_1(wL^{1/\nu})$ for which \eqref{eqScal} implies the scaling form $X_{\mathrm{m}}(W,L)=\psi_0(wL^{1/\nu})$ where $\psi_0$ is an unknown function. 
To achieve finite size collapses of the data, we minimize the following cost function 
\begin{equation}
\mathcal{F}_X = \frac{\sum_j \vert X_{j+1}-X_j\vert}{\max\lbrace X_j\rbrace-\min\lbrace X_j\rbrace}-1,
\end{equation}
(with $X_j\equiv X_{\mathrm{m}}(W_j,L_j)$ sorted according to the value of $wL^{1/\nu}$ \cite{Suntajs20}) by performing an optimization with respect to $y$, $a_0$, $a_1$ and keeping $\nu \in [0.3,3]$, $W_C \in[2.5,5]$ fixed. 

The collapses for the gap ratio $X=\overline r$ yield $\mathcal{F}_{\overline r}$ shown in Fig.~\ref{figCOMB2}(a). A wide minimum of $\mathcal{F}_{\overline r}$ in the direction $\nu \sim W_C$ shows that the FSS analysis alone is insufficient to determine the values of the critical parameters $\nu$ and $W_C$.
Assuming additionally that $W_C\approx W_{\infty}=3.97\pm0.03$, we find $\nu=1.9\pm0.1$.
We would like to note here, that the error bar of $W_{\infty}$ is associated with uncertainties of the coefficients in the assumed fitting of $W^*_{\overline r}(L)$ by a second order polynomial in $1/L$. The obtained value of $\nu$ suggests that $W_{\infty}$ is a reasonable candidate for the critical disorder strength $W_C$ of MBL transition. However, we cannot prove that the assumption about the scaling form of $W^*_{\overline r}(L)$ is valid. Thus, our numerical results are insufficient to estimate with what accuracy $W_{\infty}$ approximates the critical disorder strength $W_C$ for MBL transition in KIM.

The contours $\mathcal{F}_{\overline r}=\frac{4}{3}\mathcal{F}^{min}_{\overline r}$, which encompass the broad minimum of the cost function, shift and elongate when the system sizes considered in the collapse increase from $L=10$$-$$14$ to $L=14$$-$$18$. This highlights the importance of finite size effects and  demonstrates qualitative changes in the behavior of $\overline r$ when the system size is increased beyond $L=14$. Analogous FSS analysis performed for the Schmidt gap $\overline \Delta$, finds a much better stability of the results with respect to the system size $L$, as exhibited by $\mathcal{F}_{\overline \Delta}$ presented in Fig.~\ref{figCOMB2}(b). A similar conclusion was obtained for the XXZ spin chain \cite{Gray18}. Despite the apparent correlation between $\nu$ and $W_C$, the minimum of $\mathcal{F}_{\overline \Delta}$ is narrower, consistent with $\nu=2\pm0.5$ and $W_C = 4.1\pm0.5$. Assuming $W_C \approx W_{\infty}$, one gets $\nu=1.95\pm0.1$. We perform similar collapses for the rescaled QMI $\overline i_2$ and the spin stiffness $\overline C$. The results, summarized in Fig.~\ref{figCOMB2}(c), display the correlation $\nu \sim W_C$  for all quantities considered. The intersection of all of the contours for $W_C \approx W_{\infty}$ yields $\nu = 2\pm 0.1$ for which we obtain data collapses 
shown in Fig.~\ref{figCOMB2}(d),(e). Notably, we find that $\overline r_{\mathrm{m}} \approx \overline r_{PS} $ at the MBL transition. See App.~\ref{app:fss} for further details on the FSS analysis. 

\section{Discussion}

The premises suggesting that the ergodic-MBL crossover observed in numerical data for KIM gives rise to an MBL transition in the thermodynamic limit may be compared with features of the crossover between delocalized and localized regimes of Anderson model on random regular graphs (RRG) \cite{Abou73, Mirlin91, Evers08}. The crossover in the latter model shares similarities with the ergodic-MBL crossover \cite{Tikhonov16}, but the critical disorder strength for the Anderson transition on RRG can be accurately determined.
Investigation of Anderson model on RRG of size $\mathcal N=2^L$ and varying connectivity \cite{Sierant22} shows that: i) the boundary of the delocalized regime, $W^T_{\overline r}(L)$, follows a linear scaling with $L$ that is replaced by a weaker, sub-linear, growth at $L\approx 13$; ii) the length scale at which the linear growth of $W^T_{\overline r}(L)$ crosses with the extrapolated scaling of the crossing point $W^*_{\overline r}(L)$ is $L^{\mathrm{RRG}}_0 \approx 25$; iii) extrapolation of the crossing point $W^*_{\overline r}(L)$ to $L \to \infty$ reproduces the exactly known critical disorder strength \cite{Parisi19, Tikhonov19} with accuracy to a few percent. All these observations are in line with the findings presented in this work for KIM and support the interpretation of the results as indicating the presence of a transition to an MBL phase at the critical disorder strength $W_C \approx W_{\infty}$.

Examination of results for various parametrizations of the KIM, as well as for other Floquet models shows the robustness of the observed scalings of $W^T_{X}(L)$ and $W^*_{X}(L)$. The influence of the symmetry of the system on the ergodicity breaking is an important aspect of our results. The phenomenon of MBL does not occur in disordered spin chains with non-abelian $SU(2)$ symmetry. Instead, one observes a broad non-ergodic regime in which the ergodicity is restored only beyond certain system size \cite{Protopopov20}. One could then intuitively expect that the absence of $U(1)$ and time translation symmetries will additionally stabilize the MBL regime in KIM in comparison to the disordered XXZ spin chain. Our results indeed confirm this intuition as $L^{\mathrm{KIM}}_0$ is significantly smaller than $L^{\mathrm{XXZ}}_0$. However, the contributions of the two symmetries to this effect are much different. Our comparison of KIM with the Floquet models $U_{F,1}$, $U_{F,2}$, $U_{F,3}$ shows that the presence of the abelian symmetries such as $Z_2$ or $U(1)$ has a minor impact on the finite size effects at the MBL crossover. The major difference between  $L^{\mathrm{KIM}}_0$ and $L^{\mathrm{XXZ}}_0$ (or $L^{\mathrm{TFIM}}_0$) can be attributed to the presence or absence of the time translation symmetry in those models. Finally, according to intuitive expectations, the comparison of KIM with the Floquet models $U_{F,3}$ and $U_{F,4}$ shows that the characteristic length scale $L_0$ is quickly increasing with the range of operators used to construct the model.

\section{Conclusions}
We examined the ergodic-MBL crossover in disordered Floquet models by investigating the boundary of the ergodic regime $W^T_{X}(L)$ and the crossing point $W^*_{X}(L)$ that estimates the position of a putative transition to MBL phase. Focusing on disordered KIM, we have shown that the dependence of $W^T_{X}(L)$ and $W^*_{X}(L)$ on the system size, $L$, allows one to estimate a length scale, $L^{\mathrm{KIM} }_0$\, which quantifies the strength of finite size effects at the MBL crossover. 
We found that $L^{\mathrm{KIM} }_0\approx28$ for KIM is considerably smaller than the corresponding length scale for disordered XXZ model $L^{\mathrm{XXZ} }_0\approx 50$ \cite{Sierant20p, Panda19, Suntajs20}. This indicates that finite size effects at ergodic-MBL crossover in the former model are less severe than in the latter and  allows us to observe premises of a transition to MBL phase along the whole ergodic-MBL crossover in KIM. A linear with $L$ increase of $W^T_{X}(L)$ is replaced by a sub-linear growth at $L \geq 15$, consistent with a transition to MBL phase at a sufficiently strong disorder. The crossing points $W^*_{X}(L)$ of gap ratio ($X=\overline r$), rescaled entanglement entropy ($X=\overline s$), rescaled QMI ($X=\overline i_2$) are well approximated by polynomials in $1/L$ 
which, upon extrapolation to $L \to \infty$ limit, consistently predict an ergodic-MBL transition in KIM at $W_C \approx 4$. 
We note that finite system size effects of similar type \cite{Sierant22} are found for the Anderson localization transition on random regular graphs, a phenomenon that occurs at an exactly known critical disorder strength \cite{Tikhonov19, Parisi19}. 
Assuming a power-law divergence of the correlation length at the transition in KIM, we have shown that the estimated value of $W_C \approx 4$ is consistent with the correlation length exponent $\nu \approx 2$ fulfilling the Harris criterion \cite{Harris74, Chayes86, Chandran15a}. Considering various parametrizations of KIM as well as other disordered Floquet systems, we demonstrated the robustness of our conclusion that the finite size effects at the MBL crossover in Floquet systems are less severe than in the disordered spin chains typically considered in the context of MBL.

Our results provide numerical arguments in favor of the presence of an MBL transition in a disordered 1D quantum many-body system system. This is of particular importance in view of the recent controversies around the MBL transition in the disordered XXZ spin chain. The latter model, in contrast to KIM, possesses the time translation symmetry, which we have identified as the main factor enhancing the finite size effects at the MBL crossover in the disordered XXZ spin chain. In that sense, our findings support the intuition that the higher the symmetry of the model, the weaker the signatures of MBL. Additionally, due to the lack of $U(1)$ symmetry, the arguments of \cite{Kiefer20, Kiefer21} against the stability of MBL do not apply to KIM. The investigated ergodicity breaking in KIM is an example of MBL in Floquet systems that underlies the stability of Floquet time crystals \cite{Choi17, Bordia17, Ippoliti21, Mi22} and Floquet insulators \cite{Rudner20} by providing a mechanism to completely eliminate the heating due to periodic driving of the system.

\begin{acknowledgments}
 PS acknowledges discussions with D. Luitz at the early stages of this work. We acknowledge the support of PL-Grid Infrastructure. PS and ML acknowledge the support of ERC AdG NOQIA; Ministerio de Ciencia y Innovation Agencia Estatal de Investigaciones (PGC2018-097027-B-I00/10.13039/501100011033, CEX2019-000910-S/10.13039/501100011033, Plan National FIDEUA PID2019-106901GB-I00, FPI, QUANTERA MAQS PCI2019-111828-2, QUANTERA DYNAMITE PCI2022-132919, Proyectos de I+D+I “Retos Colaboración” QUSPIN RTC2019-007196-7); MICIIN with funding from European Union NextGenerationEU(PRTR-C17.I1) and by Generalitat de Catalunya; Fundació Cellex; Fundació Mir-Puig; Generalitat de Catalunya (European Social Fund FEDER and CERCA program, AGAUR Grant No. 2021 SGR 01452, QuantumCAT \ U16-011424, co-funded by ERDF Operational Program of Catalonia 2014-2020); Barcelona Supercomputing Center MareNostrum (FI-2022-1-0042); EU Horizon 2020 FET-OPEN OPTOlogic (Grant No 899794); EU Horizon Europe Program (Grant Agreement 101080086 — NeQST), National Science Centre, Poland (Symfonia Grant No. 2016/20/W/ST4/00314); ICFO Internal “QuantumGaudi” project; European Union’s Horizon 2020 research and innovation program under the Marie-Skłodowska-Curie grant agreement No 101029393 (STREDCH) and No 847648 (“La Caixa” Junior Leaders fellowships ID100010434: LCF/BQ/PI19/11690013, LCF/BQ/PI20/11760031, LCF/BQ/PR20/11770012, LCF/BQ/PR21/11840013). AS acknowledges financial support from: PNRR MUR project PE0000023-NQSTI.
 Research of JZ is supported by  the National Science Centre (Poland) under grants 2019/35/B/ST2/00034, 2021/03/Y/ST2/00186 (QuantEra DYNAMITE) and 2021/43/I/ST3/01142. The support by  the Priority Research Area DigiWorld under the Strategic Programme Excellence Initiative at Jagiellonian University is also acknowledged.
Views and opinions expressed in this work are, however, those of the author(s) only and do not necessarily reflect those of the European Union, European Climate, Infrastructure and Environment Executive Agency (CINEA), nor any other granting authority.  Neither the European Union nor any granting authority can be held responsible for them. 
\end{acknowledgments}

\appendix

\begin{figure*}
 \includegraphics[width=0.95\linewidth]{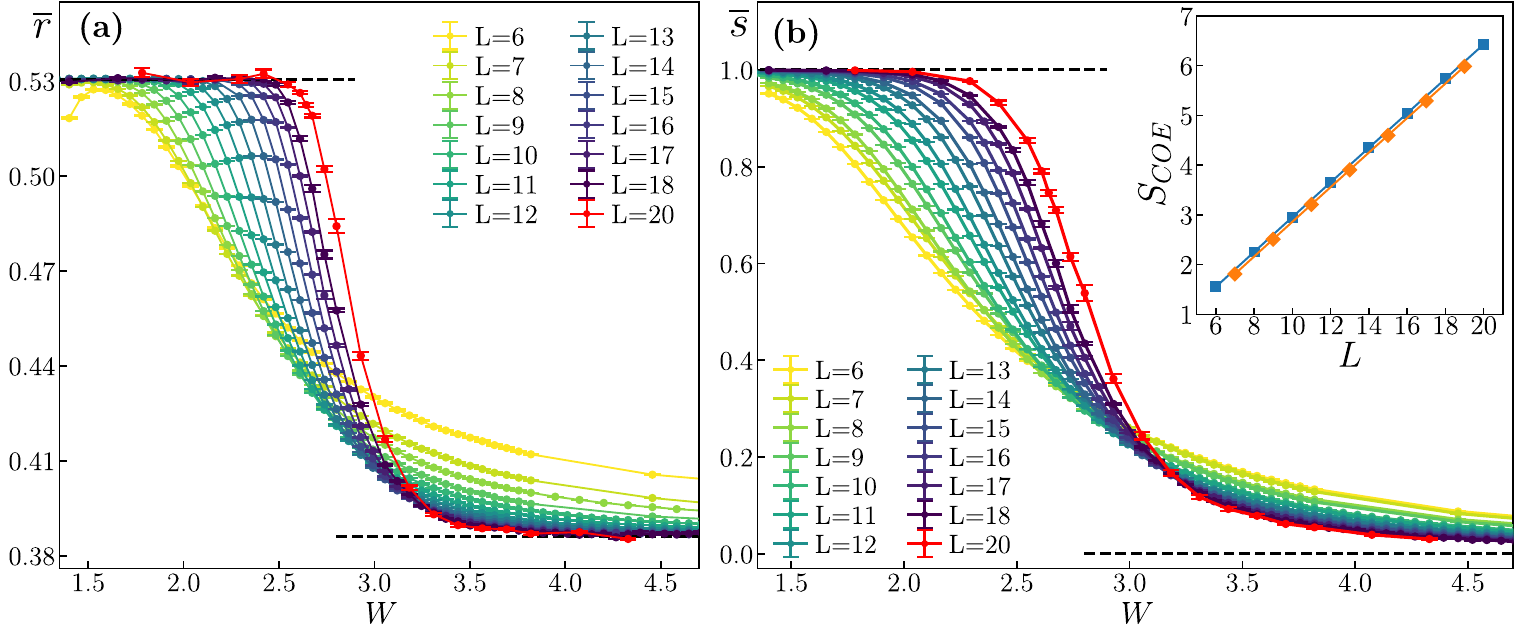} \vspace{-0.3cm}
  \caption{ The average gap ratio $\overline r$ \textbf{(a)} and the rescaled entanglement entropy $\overline s$ \textbf{(b)} as functions of disorder strength $W$ for kicked Ising model (KIM) of system size $L$. The inset in \textbf{(b)} shows the average entanglement entropy $S_{COE}$ of eigenstates of COE as function of $L$.
 }\label{supfig1}
\end{figure*}  

\section{Details of the POLFED algorithm with the geometric sum filtering}
\label{app:POLFED}

 To find eigenvectors $\ket{\psi_n}$ and the corresponding eigenphases $e^{i \phi_n }$ of the unitary operator $U_{\mathrm{KIM}}$, we employ the POLFED algorithm \cite{Sierant20p}. The algorithm is based on a block Lanczos iteration \cite{Lanczos50, Cullum74, Saad80} performed for a polynomial $g_K(U_{\mathrm{KIM}})$ of order $K$ of the matrix $U_{\mathrm{KIM}}$ (see \cite{Bekas08, Fang12, Pieper16, Guan21} for similar techniques). The matrix $g_K(U_{\mathrm{KIM}})$ has the same eigenvectors $\ket{\psi_n}$ as $U_{\mathrm{KIM}}$, but its eigenvalues are equal to $g_K(e^{i \phi_n })$. The idea of the approach is to use the polynomial $g_K$ as a \textit{spectral filter} so that its absolute value has a possibly sharp maximum for an argument $e^{i \phi_{\mathrm{tg}}}$ (where $\phi_{\mathrm{tg}}$ is a target eigenphase) at the unit circle on the complex plane. In that way, the eigenvectors $\ket{\psi_n}$ with $\phi_n$ close to  $\phi_{\mathrm{tg}}$ become eigenvectors of $g_K(U_{\mathrm{KIM}})$ to eigenvalues with dominant absolute values. 
 The Lanczos iteration converges to the eigenvectors with the largest absolute eigenvalues, which allows us to compute the eigenvectors $\ket{\psi_n}$ with $\phi_n$ close to  $\phi_{\mathrm{tg}}$.
 
 A polynomial which can be effectively used as the spectral filter for unitary operators was proposed in \cite{Luitz21}, and is simply a geometric sum:
 \begin{eqnarray}
  g_K(U_{\mathrm{KIM}}) = \sum_{m=0}^K e^{-i m \phi_{\mathrm {tg} }} U_{\mathrm{KIM}}^m.
  \label{eqsup1}
 \end{eqnarray}
The order of the polynomial $K$ is fixed by the number of requested eigenvectors $N_{\mathrm{ev}}$ and the Hilbert space dimension $\mathcal N=2^L$ as 
 \begin{eqnarray}
K=f \frac{ \mathcal  N}{N_{\mathrm{ev}}}  
\label{eqsup2}
 \end{eqnarray}
 where the factor $f=1.46$ was obtained from an optimization of the performance of the algorithm. For that choice, the algorithm converges to approximately $ N_{\mathrm{ev}}$ eigenvectors after $ \alpha N_{\mathrm{ev}}$ steps of the Lanczos iteration, where $\alpha \approx 2.1$. Each step of the Lanczos iteration involves a single multiplication of a vector by the polynomial $g_K(U_{\mathrm{KIM}})$ which reduces to $K$ multiplications of the vector by $U_{\mathrm{KIM}}$ and basic linear algebra operations. 
 Thus, the total computation cost is proportional to $\alpha N_{\mathrm{ev}} K V+ R$ where $R$ is the cost of the reorthogonalization of the vectors during the Lanczos iteration and $V$ is the cost of the single matrix vector multiplication. We employ the full reorthogonalization scheme, hence, it costs scales as $R\sim N_{\mathrm{ev}}^2 \mathcal N$. Since $V \sim L \mathcal N$ for $U_{\mathrm{KIM}}$ (as we argue below), the contribution $\alpha N_{\mathrm{ev}} K V = \alpha f L \mathcal N^2$ dominates the total computation time. Notably, this contribution is independent of the number of requested eigenvalues $N_{\mathrm{ev}}$. Hence, we can increase $N_{\mathrm{ev}}$ without a significant increase in the total computation time up to a point at which the reorthogonalization cost $R$ starts to be comparable with $\alpha f L \mathcal N^2$. This, together with considerations about memory usage (which is proportional to $N_{\mathrm{ev}} \mathcal N$) lead us to consider 
 $N_{\mathrm{ev}} = \mathrm{min}\{ 2^L /10, 1000\}$.
 
 Once the Lanczos iteration for $g_K(U_{\mathrm{KIM}})$ converges to vectors $\ket{u_i}$, we calculate the residual norms $\epsilon_i = || U_{\mathrm{KIM}} \ket{u_i} - \bra{u_i} U_{\mathrm{KIM}} \ket{u_i} u_i ||$. Even though the order $K$ of the polynomial \eqref{eqsup1} may reach few thousands for the largest considered system sizes, we find consistently that the algorithm calculates the eigenvectors of $U_{\mathrm{KIM}}$ with a high numerical accuracy and the residual error norm $ \epsilon_i < 10^{-14}$. Also, the algorithm calculates eigenvectors to all consecutive eigenphases in the vicinity of the target eigenphase $\phi_{\mathrm{tg}}=0$ so that the gap ratios $r_n$ (which are determined by three consecutive eigenphases) can be calculated without problems.
 
The computation time of the POLFED algorithm is dominated by the multiple multiplications of vectors by the matrix $U_{\mathrm{KIM}}$. To perform a single matrix vector multiplication we note that 
\begin{eqnarray}
 U_{\mathrm{KIM}}= e^{-i g \sum_{j=1}^L \sigma^x_j} e^{ -i \sum_{j=1}^L (J \sigma^z_j \sigma^z_{j+1}  +  h_j \sigma^z_j ) }
 \label{eqsup3}
\end{eqnarray}
is composed of two operators, the first diagonal in the  eigenbasis of $\sigma^z_i$ (the Z basis) and the second diagonal in the eigenbasis of $\sigma^x_i$ (the X basis). Thus, in order to calculate $U_{\mathrm{KIM}}\ket{\psi}$, we start by expressing $\ket{\psi}$ in the Z basis, and multiply it by $e^{ -i \sum_{j=1}^L (J \sigma^z_j \sigma^z_{j+1}  +  h_j \sigma^z_j ) }$ which requires only  $\mathcal O(\mathcal N)$ operations. Subsequently, we transform the vector to the X basis, multiply it by the operator $ e^{-i g \sum_{j=1}^L \sigma^x_j}$ diagonal in X basis, and finally we transform the vector back to the Z basis. To transform the vector between the bases, we employ a fast Hadamard transform \cite{Fino76, Arndt11} which requires $\mathcal O(\mathcal N \log \mathcal N)$ operations. The described procedure of multiplication by $U_{\mathrm{KIM}}$ is central for the efficiency of the POLFED approach described here, and also simplifies investigations of quantum dynamics in Floquet models \cite{Prosen99,Lezama19}.

\section{Analysis of statistical uncertainties of results}
\label{app:error}

In our analysis of the ergodic-MBL crossover we fix the disorder strength $W$ and consider quantities averaged over $N_{\mathrm{ev} }$ eigenstates/eigenvalues of the Floquet operator (or Hamiltonian in the case of TFIM) and over $N_{\mathrm{ dis} }$ disorder realizations. It has been observed that fluctuations of the rescaled entanglement entropy \cite{Khemani17a} or of the average gap ratio \cite{Sierant19b} between different disorder realizations are enhanced in the vicinity of the ergodic-MBL crossover when the system size $L$ increases. Hence, when $N_{\mathrm{ev}}$ is fixed, both $\overline r$  and $\overline s$ are not self-averaging \cite{Schiulaz19b, TorresHerrera20} (other quantities considered by us share the same problem). Assume that we fix $N_{ev}$ and calculate $r_S$, the average value of the gap ratio for a single disorder realization. The lack of self-averaging implies that a variance $\braket{ (r_S - \overline r)^2 }$, where $\braket{.}$ denotes average over disorder samples at given $W$, is not decreasing (and can be even increasing) with system size $L$. At the same time, the exponential increase of the Hilbert space dimension with system size forces us to consider smaller number of disorder realizations $N_{\mathrm{dis}}$ with increasing $L$. 

We employ the following procedure in order to estimate the statistical uncertainties of the obtained results.  For each disorder sample, we compute the average value of quantity $X_S$ (which may be the gap ratio, rescaled entanglement entropy, Schmidt gap or spin stiffness). Then, the resulting  statistical uncertainty is
\begin{equation}
\sigma_X = \frac{ \left( \braket{ (X_S - \overline X)^2 } \right)^{1/2} }{N_{\mathrm{dis}}^{1/2} },
\label{eq:err}
\end{equation}
where $\overline X = \braket{X_S}$. This procedure assumes that the values of $X_S$ for different disorder samples are uncorrelated, as reflected by $N_{\mathrm{dis}}^{1/2}$ in the denominator of \eqref{eq:err}. To test this procedure, we assumed a hypothesis that $\overline r$ as a function of $W$ at a fixed finite system size $L$ can be described, in a certain interval of $W$, by a polynomial of a small order in $W$. Performing fitting with polynomials of degree $5$ to $~16$ points in the vicinity of the crossing points for KIM data at $L=12, 14, 16, 18, 20$, we have obtained values of $\chi^2$ per degree of freedom between $0.8$ and $1.7$ suggesting that our analysis well estimates the statistical uncertainty of the calculated quantities. 

\begin{figure}
 \includegraphics[width=0.95\linewidth]{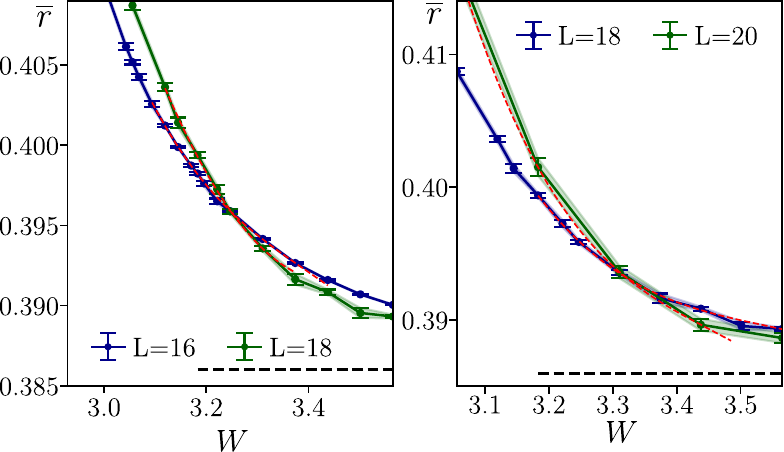} \vspace{-0.3cm}
  \caption{Extraction of $W^*_{\overline r}(L)$ for KIM with $g=J=1/W$ for $L=16,18$ (\textbf{a}) and $L=18,20$ (\textbf{b}). The shaded regions correspond to the estimated statistical uncertainties of $\overline r$, see \eqref{eq:err}. The $\overline r(W)$ curves are locally fitted with a polynomials of order $2$ or $3$ (denoted by the red dashed-lines) and the value of $W^*_{\overline r}(L)$ is extracted as the crossing point of the respective polynomials.
 }\label{supfig:err}
\end{figure}

Importantly, the decrease of $N_{\mathrm{dis}}$ at the largest system sizes available to us, yields larger statistical uncertainties of the obtained values of $W^T_{X}(L)$ and $W^*_X(L)$. This is illustrated in Fig.~\ref{supfig:err}, where the vicinity of the crossing points for $L=16,18$ and $L=18,20$ is shown. While the shaded areas corresponding to the uncertainty of $\overline r$ are wider for larger $L$, the obtained numbers of disorder realization allows us to relatively accurately extract the value of $W^*_{\overline r}(L)$. Similar applies to $X=\overline s, \overline i_2$. Finally, we note that the sample-to-sample fluctuations of $X_S$ close to the ergodic region are significantly weaker than those close to the crossing point. This leads to a smaller uncertainty of the extracted values of $W^T_{\overline r}(L)$ as compared to $W^*_{\overline r}(L)$.

\begin{figure}
 \includegraphics[width=0.95\linewidth]{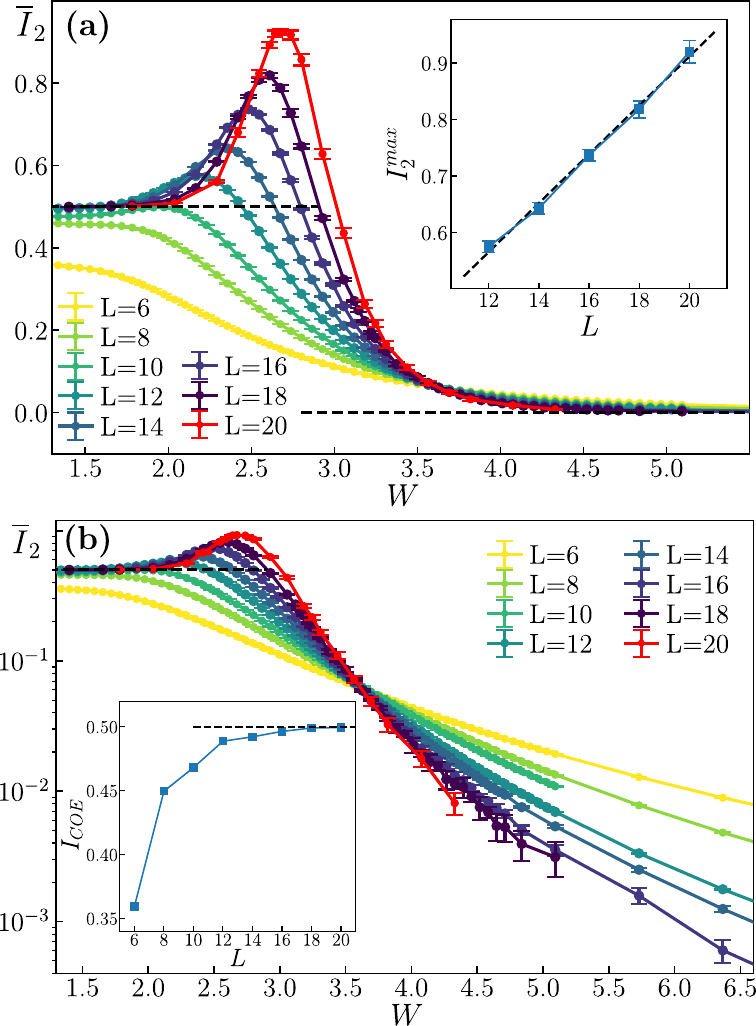} \vspace{-0.3cm}
  \caption{The average quantum mutual information (QMI) $\overline I_2$ as function of disorder strength $W$ for KIM of system size $L$; panel \textbf{(a)} - linear vertical scale, panel \textbf{(b)} - logarithmic vertical scale.  The inset in \textbf{(a)} shows $\overline I^{max}_2$, the maximum of $\overline I_2$, as a function of system size $L$. 
  The inset in \textbf{(b)} shows the average QMI $I_{COE}$ of eigenstates of COE as function of $L$.
 }\label{supfig2}
\end{figure}

\begin{figure*}
 \includegraphics[width=0.9\linewidth]{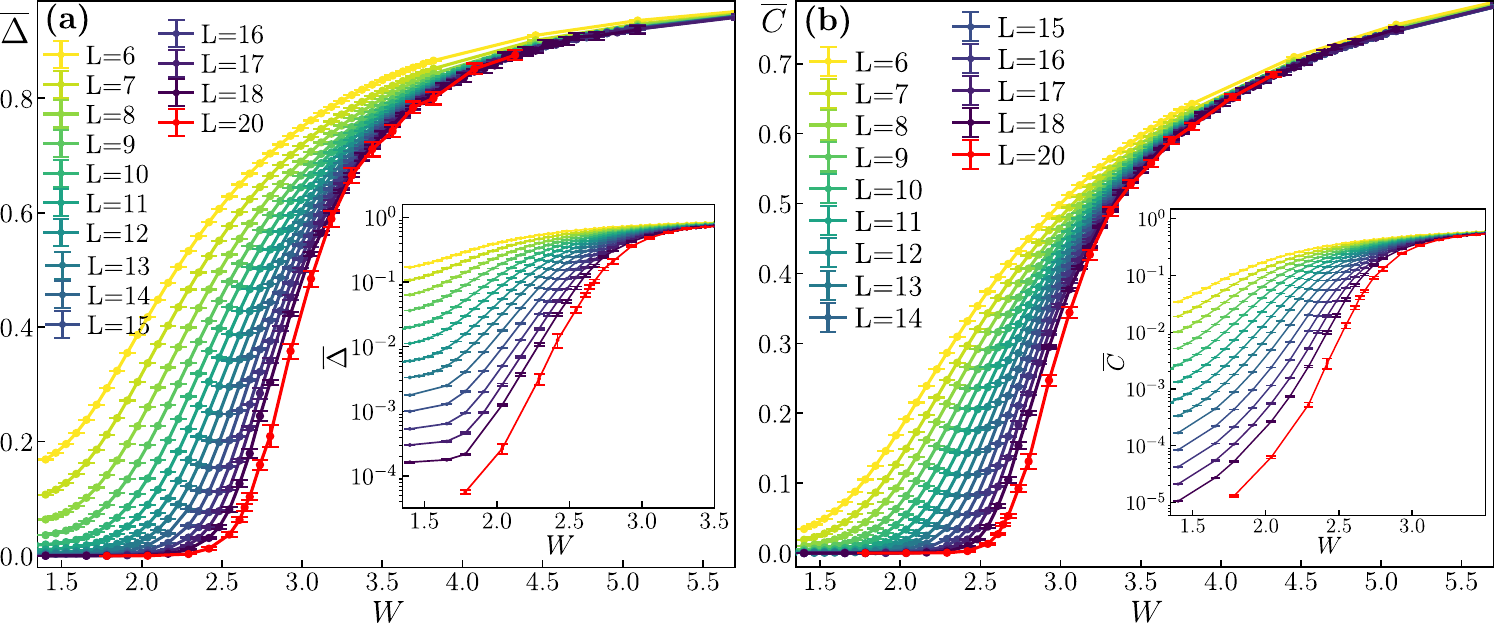} \vspace{-0.3cm}
  \caption{ The average Schmidt gap $\overline \Delta$ \textbf{(a)} and spin stiffness 
  $\overline C$ \textbf{(b)} as functions of disorder strength $W$ for KIM of system size $L$. The insets show the same, but using a logarithmic vertical axis.
 }\label{supfig3}
\end{figure*}  

 \section{Ergodic-MBL crossover in KIM}
 \label{app:addNUMkim}
 A complete set of data for the average gap ratio $\overline r$, used in the determination of disorder strengths $W^T_{\overline r}(L)$ and $W^*_{\overline r}(L)$, is shown in Fig.~\ref{supfig1}(a). For KIM defined on chain of length $L\geq 6$, we observe a crossover between the ergodic regime $\overline r \approx r_{COE} \approx 0.53$ and MBL regime with $\overline r = \overline r_{PS}\approx 0.386$, which is a value for Poissonian level statistics that emerges due to the presence of local integrals of motion in the system.
 
 The ergodic-MBL crossover looks qualitatively similar from the perspective of the rescaled entanglement entropy $\overline s = \overline S/S_{COE}$ which changes from  $1$ to $0$ between the ergodic and MBL regimes, see Fig.~\ref{supfig1}(b). In the ergodic regime the average entanglement entropy $\overline S$ is well approximated by the entanglement entropy $S_{COE}$ of eigenstates of COE, shown in the inset in Fig.~\ref{supfig1}(b). The entanglement entropy $S_{COE}$ increases according to a volume-law, i.e. proportionally to system size $L$. Linear fits yield $S_{COE}=aL+b$ with:  
 $a$$=$$0.350$ ($a$$=$$0.347$) and $b$$=$$-0.551$ ($b$$=$$-0.509$) for even system sizes $L$$=$$6,8,10,12$ ($L$$=$$14,16,18,20$) and $a$$=$$0.349$ ($a$$=$$0.347$) and $b$$=$$-0.634$ ($b$$=$$-0.603$) for system sizes $L$$=$$7,9,11,13$ ($L$$=$$15,17,19$) showing that the coefficient $a$ approaches the expected value $\ln(2)/2\approx 0.34657$ with increasing system size \cite{Vidmar17}.

The average QMI $\overline I_2$, shown as a function of disorder strength $W$ in Fig.~\ref{supfig2}, admits a maximum at disorder strength $W^m_{\overline i_2}(L)$ for system size $L$. The value  $\overline I^{max}_2$ of the average QMI at the maximum is shown in the inset in Fig.~\ref{supfig2}(a) as a function of $L$. We observe that $\overline I^{max}_2$ scales approximately linearly with the system size $L$.  The inset in Fig.~\ref{supfig2}(b) shows that the average QMI of COE eigenstates, $I_{COE}$, saturates with the increase of $L$ to a system size independent value $I_{COE}\approx 0.5$. As Fig.~\ref{supfig2}(b) shows, the average QMI $\overline I_2$ decreases approximately exponentially with disorder $W$ as well as with the system size $L$ in the MBL regime. 

The behavior of the Schmidt gap $\overline \Delta$ and spin stiffness $\overline C$ across the ergodic-MBL crossover is shown in Fig.~\ref{supfig3}. In contrast to $\overline r$, $\overline s$ and $\overline i_2$, the Schmidt gap and spin stiffness decrease monotonically with increasing system size (consequently, there are no crossing points that could be used to perform an analysis with disorder strength $W^*_X(L)$ for those quantities). The rate of the decrease is, however, markedly different in the ergodic and MBL regimes. In the former, $\overline \Delta$ and $\overline C$ decrease approximately exponentially with system size $L$ (as demonstrated by the insets in Fig.~\ref{supfig3}). In the latter regime, the decrease of the Schmidt gap and spin stiffness with $L$ is much slower and at $W\gtrsim 4$, $L\geq10$ the value of $\overline \Delta$ and $\overline C$ appears to be independent, within the estimated error bars, of the system size $L$,  consistently with the prediction that at $W \geq W_{\infty}\approx 4$ the KIM is in the MBL phase.

\begin{figure}
 \includegraphics[width=0.85\linewidth]{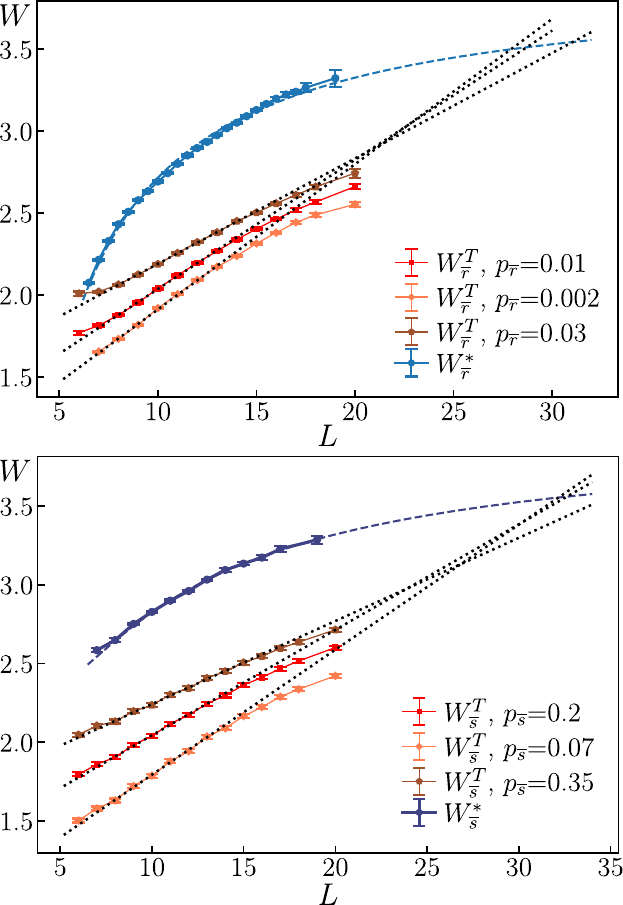} \vspace{-0.3cm}
  \caption{The disorder strength $W^T_X(L)$ for various choices of the threshold $p_X$ compared with the crossing point $W^*_X(L)$. Panel \textbf{(a)} shows results for the average gap ratio $X=\overline r$, panel \textbf{(b)} presents results for the rescaled entanglement entropy $X=\overline s$. The dashed lines show an extrapolation of $W^*_X(L)$ with a second order polynomial in $1/L$, whereas the dotted lines denote first order polynomial in $L$ fits in the regime of linear growth of $W^T_X(L)$.
 }\label{supfig4}
\end{figure}  

\section{The robustness of scaling of  $W^T_X(L)$ with system size}
\label{app:px}

In this section we analyze the robustness of the system size dependence of the disorder strength $W^T_X(L)$ at which the quantity $X$ deviates from its ergodic value by a small parameter $p_X$. Fig.~\ref{supfig4} shows $W^T_X(L)$ for various choices of $p_X$ for the gap ratio $X=\overline r$ and for the rescaled entanglement entropy $X=\overline s$.

In Fig.~\ref{supfig4}(a) we observe a regime of linear increase of $W^T_{\overline r}(L)$ for 
$7 \leq L \leq 14 $ and a deviation from this linear scaling at $L\geq 15$ for the considered values of $ p_{\overline r}\in [0.002, 0.03]$. This confirms that the conclusions about system size scaling of $W^T_{\overline r}(L)$ reported in the main text are robust with respect to changes of $ p_{\overline r}$. The length scale $L^{\mathrm{KIM} }_0$ is mildly dependent on $p_{\overline r}$, and it does not exceed $30$ lattice sites for the considered interval of $p_{\overline r}$.

Analysis of the rescaled entanglement entropy yields $W^T_{\overline s}(L)$ shown in Fig.~\ref{supfig4}(b). The conclusions are the same as for $W^T_{\overline r}(L)$. There is a regime of a linear increase of $W^T_{\overline s}(L)$ with $L$ for $6 \leq L \leq 14 $ which is replaced by a sub-linear growth of $W^T_{\overline s}(L)$  for $L\geq 15$ (consistently with the presence of MBL transition at sufficiently large $W$). For $0.07 \leq p_{\overline s}\leq 0.2$, we observe that an extrapolation of $W^T_{\overline s}(L)$ yields 
$\tilde L^{\mathrm{KIM}}_0 \approx 32$ which is close to the length scale $L^{\mathrm{KIM} }_0$ obtained from the extrapolation of the linear scaling of $W^T_{\overline r}(L)$.

\begin{figure*}
 \includegraphics[width=0.99\linewidth]{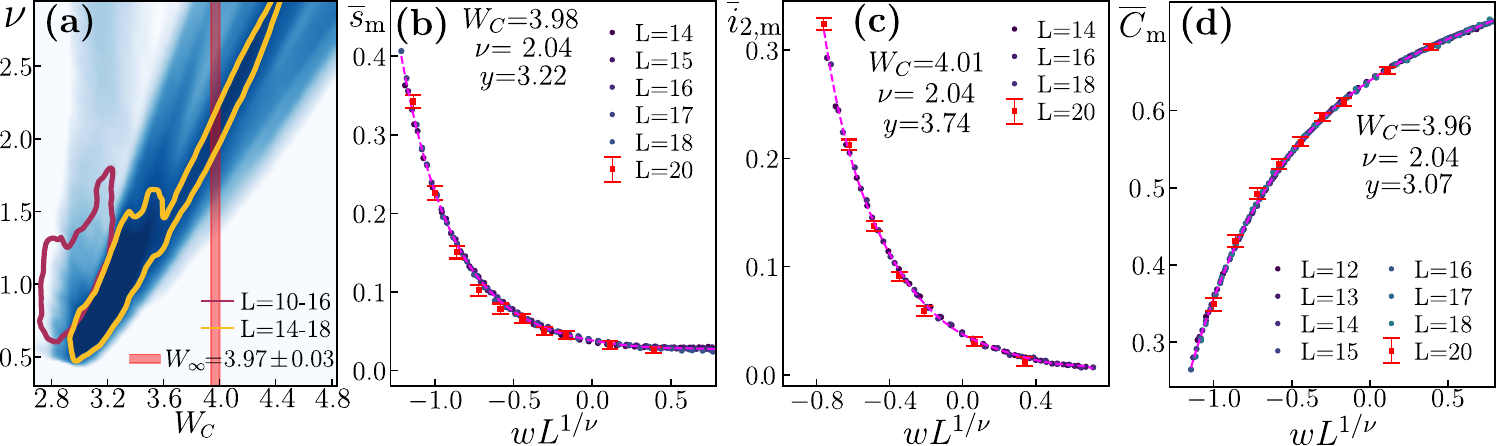} \vspace{-0.3cm}
  \caption{ Supplementary data for finite size scaling analysis of ergodic-MBL crossover in KIM. 
  Cost function $\mathcal F_{\overline s}$ for the collapse of the rescaled entanglement entropy $\overline s$ is color coded for fixed $\nu$, $W_C$ in panel \textbf{(a)} (system sizes considered in the collapse $L=14$$-$$18$).
  The contours highlight the change in the cost function system size by 
  encompassing region of $\nu$ and $W_C$ for which $\mathcal F_{ \overline s} < 2 \mathcal F^{\mathrm{min}}_{\overline s}$ where $F^{\mathrm{min}}_{\overline s}$ is the minimum of $\mathcal F_{\overline s}$.
 Collapses for the rescaled entanglement entropy $\overline s_{\mathrm{m}}$, rescaled QMI $\overline i_{2,\mathrm{m}}$ and spin stiffness $\overline C_{\mathrm{m}}$ shown in \textbf{(b)}, \textbf{(c)}, \textbf{(c)}; $w=(W-W_C)/W_C$ is the dimensionless distance from the critical point and the plots show the quantities with subtracted sub-leading correction to the scaling $X_{\mathrm{m}}\equiv X-L^{-y}\psi_1(wL^{1/\nu})$.
 }\label{supfig5}
\end{figure*}  

\section{Additional data for finite size scaling analysis}
\label{app:fss}

In this section we provide additional data for the finite size scaling analysis at the ergodic-MBL crossover in KIM. Fig.~\ref{supfig5}(a) shows the cost function $\mathcal{F}_{\overline s}$ for the collapse of rescaled entanglement entropy $\overline s$. The conclusions are similar as for the gap ratio collapses reported in the main text. At sufficiently large system sizes ($L=14-18$), there appears a wide minimum of the cost function. This minimum is consistent with a broad interval of critical disorder strength $W_C$ and exponent $\nu$. Assuming, additionally, that $W_C\approx W_{\infty} \approx 4$, one obtains the power-law exponent $\nu\approx 2$ that is consistent with the Harris criterion for 1D disordered systems.  The corresponding collapse of the data for the rescaled entanglement entropy is shown in Fig.~\ref{supfig5}(b). Collapses for $\nu \approx 2$ and $W_C\approx 4$ for the rescaled QMI $\overline i_2$ and for the spin stiffness $\overline C$ are shown in Fig.~\ref{supfig5}(c) and (d). Interestingly, the collapse of the gap ratio $\overline r$ shown in the main text predicts that $\overline r_{\mathrm m}$ is equal to $\overline r_{PS} \approx 0.386$ characteristic for a localized system at the critical point $W=W_C$. At the same time, the values of $\overline s_{\mathrm{m}}$ and $\overline i_{2,\mathrm{m}}$ seem to be not vanishing at $W=W_C$ despite being significantly smaller than their respective ergodic values.

\normalem

\begin{thebibliography}{168}
\bibitem [{foo({\natexlab{b}})}]{footnoteWT}%
  \BibitemOpen
  \href@noop {} {\emph {\bibinfo {title} {\normalfont{ $
  \lim_{L\to\infty}W^T_{\overline r}(L) \approx 3.97 < W_{\infty}$, but since
  $W^T_{\overline r}$ is the most distant point of the crossover from the
  critical point, we regard this value as a lower bound for the critical
  disorder strength }}}}\BibitemShut {NoStop}%
\end{thebibliography}

%





\end{document}